\documentclass[%
reprint,
superscriptaddress,
amsmath,amssymb,
aps,
pra,
]{revtex4-2}

\usepackage{graphicx}
\usepackage{braket}
\usepackage[hidelinks]{hyperref}
\usepackage{booktabs, multirow}
\usepackage{amsthm}

\newtheorem{definition}{Definition}[]

\begin{document}

\title{High-order splitting of non-unitary operators on quantum computers}

\author{Peter Brearley}
\email{peter.brearley@manchester.ac.uk}
\affiliation{Centre for Quantum Science and Engineering, University of Manchester, UK}
\affiliation{Department of Mechanical and Aerospace Engineering, University of Manchester, UK}
\author{Philipp Pfeffer}
\affiliation{Institute of Thermodynamics and Fluid Mechanics, Ilmenau University of Technology, Germany}

\begin{abstract}
    Dissipation and irreversibility are central to many physical systems, yet they lead to non-unitary dynamics that are challenging to realise on quantum processors. High-order operator splitting is an attractive approach for simulating unitary dynamics, yet conventional product formulas introduce negative time steps at high orders that are ill-conditioned for dissipative dynamics. We show how block encodings of complex-coefficient product formulas can be constructed by a sequence of simple Hamiltonian evolutions in real and imaginary time with high-order accuracy. The unitary substages use positive real coefficients, while the dissipative substages use complex coefficients with positive real parts; the real parts preserve the contractive evolution, and the imaginary parts are additional unitary evolutions. We use known formulas of orders 4 and 6 and give a new symmetric formula of order 8 in this restricted class. We apply the approach to the classical problem of lossy mechanical wave propagation in statevector simulations and on a trapped-ion quantum processor. Orders 4 and 6 require comparable CNOT gates to orders 1 and 2 at large error tolerances, but quickly become more efficient as the tolerance is reduced. On the quantum hardware, a single step of order 4 achieves lower error than order 1 and comparable error to order 2, while order 6 is noise-limited. Our results show that operator splitting can combine practical circuit constructions with high-order convergence for non-unitary dissipative dynamics.
\end{abstract}

\maketitle

\section{Introduction}
\label{sec:introduction}

Quantum computers are naturally suited to simulating unitary dynamics because their native gate model implements the reversible time evolution of local Hamiltonians~\cite{Lloyd1996}. In contrast, many problems of practical interest in computational science are dissipative, arising from mechanisms such as friction, viscosity, diffusion, or boundary flux. Even in descriptions of quantum systems, dynamics are invariably influenced by uncontrolled environmental interactions~\cite{Kliesch2011}, motivating descriptions of dissipative open-system dynamics~\cite{Lindblad1976, Ma2024}. Developing accurate and efficient methods for simulating non-unitary, dissipative dynamics on quantum computers is therefore an important step towards supporting the adoption of quantum computing more broadly across computational science~\cite{Ding2024}. 

Operator splitting is a versatile framework for simulating Hamiltonian dynamics on quantum computers~\cite{Lloyd1996}, and high-order extensions combine this versatility with high accuracy~\cite{Childs2019, Morales2025}. Real-coefficient product formulas of orders higher than 2 necessarily contain negative coefficients~\cite{Suzuki1993, Blanes2005}, which require simulating the reverse dynamics. This is straightforward for Hamiltonian simulation, since the reverse dynamics are also unitary. For dissipative dynamics, however, the reverse evolution is numerically unstable and cannot be block-encoded into unitary operators without rescaling~\cite{Camps2024}. High-order operator splitting for the quantum simulation of dissipative dynamics is therefore yet to be demonstrated~\cite{Childs2017}, with existing applications being limited to order 2~\cite{Pfeffer2025}. 

Operator splitting is typically applied to time-evolution problems that can be written as a first-order system of ordinary differential equations (ODEs). In the linear case, this is
\begin{equation}
    \frac{\text{d}\vec{\psi}}{\text{d}t} = M\vec{\psi}
    \label{eq:ode}
\end{equation}
for a time-independent generator $M$. This system may arise by discretising linear partial differential equations (PDEs), or by linearising a non-linear system. When discretised by $N$ grid points, a PDE of order $m$ in time becomes an $mN$-dimensional system of ODEs. Efficient algorithms for integrating Eq.~\eqref{eq:ode} without operator splitting have been proposed~\cite{An2023, Jin2023}, but explicit proof-of-concept circuit constructions remain beyond the current generation of quantum processors~\cite{Hu2024}. High-order time-marching methods provide an alternative route~\cite{Ding2024}, but are high-order approximations to the complete time step, so do not benefit from implementation by a sequence of simpler, independently implementable evolutions.

To simulate Eq.~\eqref{eq:ode} by operator splitting, we consider the initial decomposition of $M$ into its Hermitian and anti-Hermitian components as $M = H_1 + iH_2$, where $H_1 = H_1^\dagger$ and $H_2 = H_2^\dagger$. Individual PDE terms often naturally separate into Hermitian and anti-Hermitian contributions, subject to boundary treatment and spatial discretisation. Even-order spatial derivatives typically contribute to the Hermitian component $H_1$, which is non-positive for contractive terms, while odd-order spatial derivatives typically contribute to the anti-Hermitian component $iH_2$, which generates unitary evolution. For systems where the dynamics do not naturally separate, these components can instead be defined as~\cite{An2023, Jin2023}
\begin{equation}
    H_1 = \frac{M+M^\dagger}{2},\qquad iH_2 = \frac{M-M^\dagger}{2}\,.
    \label{eq:hermitian_decomposition}
\end{equation}
We assume $H_1\preceq 0$, so that the real-time dissipative evolution $e^{H_1t}$ is contractive for every $t> 0$. The same decomposition extends to systems of inhomogeneous ODEs of the form $\text{d}_t\vec{\psi}=M\vec{\psi}+\vec{b}$, which can be homogenised by a simple block-matrix construction~\cite{Jin2023}, although the presence of source terms may not preserve $H_1\preceq0$.

Equation \eqref{eq:ode} has the solution $\vec{\psi}(t) = e^{Mt}\vec{\psi}(0)$, which can also be written as $\vec{\psi}(t) = e^{H_1t+iH_2t}\vec{\psi}(0)$ by Eq.~\eqref{eq:hermitian_decomposition}. The split evolution $e^{iH_2t}e^{H_1 t}$ is only equal to $e^{H_1t+iH_2t}$ if $H_1$ and $H_2$ commute, i.e.~$[H_1,H_2] = 0$. This occurs for dynamics that simultaneously diagonalise in the same basis, which often also admit analytical solutions. Therefore, operator splitting is generally required for non-trivial problems~\cite{Pfeffer2025}.

The order-1 Lie-Trotter splitting may be written as
\begin{equation}
    e^{Mt} = \left(e^{H_1 h}e^{iH_2h}\right)^{t/h} + O(t\|M\|^{2}h)\,,
    \label{eq:lie_trotter}
\end{equation}
conservatively assuming that both $\|H_1\|$ and $\|H_2\|$ are $O(\|M\|)$. The order-2 Strang splitting is then 
\begin{align}
    e^{Mt} = \left(e^{iH_2\frac{h}{2}}e^{H_1 h}e^{iH_2\frac{h}{2}}\right)^{t/h} + O(t\|M\|^3h^2)\,.
    \label{eq:strang}
\end{align}
Throughout this work, matrix norms denote the spectral norm. The overall evolution by time $t$ is discretised into $T=t/h\in \mathbb{N}$ steps of size $h$. Improving on the order-1 splitting in Eq.~\eqref{eq:lie_trotter} to the order-2 splitting in Eq.~\eqref{eq:strang} is straightforward, as it requires simulating one evolution between two half-steps of the other, and yields a substantial improvement in accuracy. The reversed Lie-Trotter composition $e^{iH_2h}e^{H_1 h}$ is likewise first order, and the complementary Strang composition $e^{H_1h/2}e^{iH_2h}e^{H_1h/2}$ is likewise second order. The ordering in Eq.~\eqref{eq:strang} may be preferable in many quantum contexts since the non-unitary evolution is confined to a single stage.

The evolutions by $e^{H_1t}$ and $e^{iH_2t}$ are usually much simpler to implement on quantum computers than $e^{Mt}$, as they represent reduced mechanisms of the dynamics. Furthermore, problems of this type have been extensively studied in the context of Hamiltonian simulation~\cite{Low2017} and imaginary time evolution~\cite{Motta2020, Gilyen2019} algorithms. If they cannot be implemented directly, then $H_1$ and $H_2$ may be further split into simpler Hamiltonians that can be efficiently simulated, provided that the dissipative subterms remain negative semidefinite. Therefore, operator splitting is a practical means of simulating dissipative dynamics on quantum computers~\cite{Pfeffer2025}, but its usefulness is severely limited by the accuracy of Eqs.~\eqref{eq:lie_trotter} and \eqref{eq:strang}, which require deep circuits to converge to within a reasonable tolerance for most practical systems.

In this work, we demonstrate a stable quantum algorithm for integrating non-unitary dynamics on quantum computers by high-order operator splitting. We consider a restricted class of complex-coefficient product formulas such that the dynamics can be expressed as a sequence of simple Hamiltonian evolutions, each corresponding to a unitary real-time evolution or a dissipative imaginary-time evolution. This gives a block encoding of the high-order product formula with no additional subnormalisation penalty from composing optimally subnormalised dissipative substages. Qualifying product formulas up to order 6 already exist~\cite{Blanes2013, Bernier2023}; we give a new qualifying product formula of order 8. 

In the next section, we present the required high-order product formulas of orders 4, 6 and 8 and discuss their implementation on a quantum computer. We then give a practical example in Section~\ref{sec:application_to_damped_waves} based on the damped wave equation, which we validate using statevector simulations before executing the corresponding circuits on the IonQ Forte 1 trapped-ion quantum processor~\cite{Chen2024}. In Section~\ref{sec:resource_requirements}, we derive the gate complexity along with logical CNOT gate counts for the damped-wave example. Finally, we discuss our conclusions and opportunities for future work in Section~\ref{sec:conclusions}.

\section{High-Order Operator Splitting}
\label{sec:high-order_operator_splitting}

\begin{table*}[t] \centering\footnotesize
\setlength{\tabcolsep}{9pt}
\caption{High-order product formulas with coefficients $a_j\in\mathbb{R}$, $b_j\in\mathbb{C}$, where $a_j\geq0$ and $\Re(b_j)>0$.}
\label{tab:splitting_coefficients}

\begin{tabular}{@{}ll@{}}
\toprule
\textbf{Product formula} & \textbf{Complex coefficient} $\boldsymbol{b_j}$ \\
\midrule

\begin{tabular}[c]{@{}l@{}}
Order 4~\cite{Castella2009} \\
$a_0,a_1,a_2,a_{3} = 1/4$
\end{tabular}
& $\begin{array}{@{}c@{\;}c@{\;}c@{\;}c@{\;}c@{\;}c@{\;}l@{}}
b_0 & = & b_4 & = & 1/10 & - & i/30 \\
b_1 & = & b_3 & = & 4/15 & + & 2i/15 \\
b_2 & = &     &   & 4/15 & - & i/5
\end{array}$ \\
\midrule

\begin{tabular}[c]{@{}l@{}}
Order 6~\cite{Blanes2013} \\
$a_0,\dots,a_{15} = 1/16$
\end{tabular}
& $\begin{array}{@{}c@{\;}c@{\;}c@{\;}c@{\;}c@{\;}c@{\;}c@{}} 
b_0 & = & b_{16} & = & 0.0246948760870180646409108649968422478386 & - & 0.0078747955629068770581715779495269421632i \\
b_1 & = & b_{15} & = & 0.0638134740213026997793663041882001469632 & + & 0.0353657610341433278046294046497147418127i \\
b_2 & = & b_{14} & = & 0.0684250940303164419703970078217446840585 & - & 0.0622622444507486769953325406444475960461i \\
b_3 & = & b_{13} & = & 0.0880477010922678376269971958694086675772 & + & 0.04547387150229870438376254918797742644469i \\
b_4 & = & b_{12} & = & 0.02368961112984706069614191247000936432533 & + & 0.00962432606408962405769803529063730666395i \\
b_5 & = & b_{11} & = & 0.04272972238677338220296430057707421855388 & - & 0.03399440392395761055408394845784435826499i \\
b_6 & = & b_{10} & = & 0.1223346863168457729604285170019625630788 & - & 0.01043585907975251066938082710059054955178i \\
b_7 & = & b_9    & = & 0.04189843282969388604353685060726223976426 & + & 0.0693624926316963842751581743071442621303i \\
b_8 &   &        & = & 0.0487328042118697081585140929349917356808 & - & 0.0905182964297247304885585385661285820513i
\end{array}$ \\
\midrule

\begin{tabular}[c]{@{}l@{}}
Order 8 (new) \\
$a_0,\dots,a_{63} = 1/64$
\end{tabular}
& See Table~\ref{tab:order8_coefficients} of the appendix
\\
\bottomrule
\end{tabular}
\end{table*}

An order-$p$ product formula $\Phi_p(t, h)$ for simulating the linear dynamics in Eq.~\eqref{eq:ode} satisfies
\begin{equation}
    e^{Mt} = \Phi_p(t, h) + O(t\|M\|^{p+1}h^p)\,,
    \label{eq:operator_splitting}
\end{equation}
comprising $q_p$ stages and $T = t/h \in \mathbb{N}$ steps such that
\begin{equation}
    \Phi_p(t, h) = \mathcal S_p(h)^T =  \left(S_{q_p-1}\cdots S_1S_0\right)^T,
    \label{eq:product_formula}
\end{equation}
where $S_j$ is a stage in the splitting step $\mathcal S_p(h)$. We utilise complex-coefficient product formulas to evolve Eq.~\eqref{eq:ode} for $M = H_1 + iH_2$ given by Eq.~\eqref{eq:hermitian_decomposition}. The unitary substages $e^{ia_jH_2h}$ are evolved by real coefficients $a_j\geq0$, and the dissipative substages $e^{H_1b_jh}$ are evolved by complex coefficients $b_j$ satisfying $\Re(b_j)>0$. Since $b_j$ is a scalar, the evolution under its real part $e^{H_1 \Re(b_j)h}$ and imaginary part $e^{iH_1\Im(b_j)h}$ commute, and can therefore be implemented individually without additional error. An individual stage in the high-order product formula $S_j(h) = V_j(h)D_j(h)$ therefore consists of a contraction $D_j$ and a unitary $V_j$, where
\begin{equation}
    D_j(h) = e^{H_1 \Re(b_j)h}, \quad V_j(h) = e^{iH_2a_jh}e^{iH_1\Im(b_j)h}.
    \label{eq:stage_definition}
\end{equation}

Each substage in Eq.~\eqref{eq:stage_definition} is typically much simpler to realise on a quantum computer than the exact evolution $e^{H_1t+iH_2t}$. If the implementation remains challenging, the Hamiltonians may be further decomposed as $H_1=\sum_k A_k$ and $H_2=\sum_k B_k$, where the individual terms can be simulated efficiently. The resulting inner splitting must match the order of the outer splitting. The unitary substages in $V_j$ may be decomposed using high-order real-coefficient product formulas~\cite{Yoshida1990}. For the contractive $D_j$, each $A_k$ must be negative semidefinite, and both coefficient families in this inner splitting may then be complex with positive real parts. Complex-coefficient splittings with this property have been developed for parabolic and analytic-semigroup evolution~\cite{Castella2009,Hansen2009,Blanes2010,Blanes2013}, with orders up to 14 obtained by recursive composition~\cite{Hansen2009}.

A product formula that satisfies the constraints of $a_j\geq0$ and $\Re(b_j)>0$ was shown by \citeauthor{Castella2009}~\cite{Castella2009}, which approximates $e^{H_1t+iH_2t}$ to order 4 using 5 stages. It uses the positive real coefficient $a_j=1/4$ across four of the stages $j=0,1,2,3$, and the terminal coefficient $a_4=0$ as required by this symmetric representation. The dissipative dynamics evolve under complex coefficients with positive real parts $b_0 = 1/10-i/30$, $b_1 = 4/15+2i/15$, $b_2 = 4/15-i/5$, $b_3=b_1$ and $b_4=b_0$. The dynamics can then be simulated by Eqs.~(\ref{eq:operator_splitting}--\ref{eq:stage_definition}). An order-6 product formula that satisfies the same constraints of $a_j\geq0$ and $\Re(b_j)>0$ consists of 17 stages~\cite{Blanes2013}, which is given in Table~\ref{tab:splitting_coefficients}. In the appendix, Table~\ref{tab:order8_coefficients} gives a new order-8 product formula that, to the best of our knowledge, is the highest order reported within this restricted class. The coefficients were computed numerically using arbitrary-precision Gauss-Newton iteration applied to the Lyndon-basis order conditions. The resulting coefficients were verified against the full word-basis order conditions to high numerical precision; the 64 most significant digits are reported in Table~\ref{tab:order8_coefficients}.

All of the reported high-order product formulas satisfy our constraints of $a_j\geq0$ and $\Re(b_j)>0$, and are symmetric, so they satisfy inversion under time-reversal $\mathcal S_p(-h) = \mathcal S_p(h)^{-1}$~\cite{Blanes2024}. Complex-coefficient formulas have also been considered for approximating unitary Hamiltonian evolution~\cite{Prosen2006, Ostmeyer2023}, but with non-unitary complex-time factors. In the present method, the complex coefficients instead multiply a negative semidefinite generator, so their real-time components remain contractive and can be block encoded. Additional qualifying formulas of orders 4 and 6 have been reported, including symmetric-conjugate schemes~\cite{Bernier2023}, which instead satisfy the conjugate transpose under time-reversal, $\mathcal S_p(-h) = \mathcal S_p(h)^{\dagger}$. This property is particularly beneficial for simulating unitary, reversible dynamics as it prevents drift in the solution norm~\cite{Bernier2023}. Since our application is intrinsically non-unitary, this advantage is less relevant. Symmetric schemes have the additional benefit that the BCH expansion of $\log \mathcal S_p(h)$ contains only odd powers of $h$, thereby eliminating the even-order conditions~\cite{Blanes2024}.

In principle, the presented technique can be applied to any product formula with coefficients satisfying $a_j\in\mathbb{R}$ and $b_j\in\mathbb{C}$ with $\Re(b_j)>0$. The coefficients used here are subject to the additional constraint of $a_j\geq0$. This is a desirable property, but it is not strictly necessary. Relaxing this constraint may ease the discovery of coefficients with order 10 and beyond, which is of theoretical but diminishing practical interest, as the required number of coefficients per step grows rapidly with the order~\cite{McLachlan1995}.

\begin{definition}[Block encoding]
    An $(n + n_{\rm a})$-qubit unitary $U_A$ is an $(\alpha,n_{\rm a},\epsilon)$-block encoding of an $n$-qubit operator $A$ if $\| A - \alpha(\bra{0}^{\otimes n_{\rm a}} \otimes I)U_A(\ket{0}^{\otimes n_{\rm a}} \otimes I) \| \le \epsilon$.
    \label{def:block_encoding}
\end{definition}
The quantity $\alpha\geq\|A\|-\epsilon$ is known as the subnormalisation. The objective on a quantum computer is to prepare an $(\alpha, n_{\rm a}, \epsilon)$-block encoding $U_A$ of $A = e^{Mt}$. We do so by preparing an exact $(\alpha_\Phi,n_{\rm a}, 0)$-block encoding $U_\Phi$ of the high-order product formula $\Phi_p(t, h)$. Assume that every unitary substage $V_j$ is implemented exactly, and that $U_{D_j}$ is an exact $(\alpha_j,n_j,0)$-block encoding of $D_j$. Then, $U_{S_j} = (I\otimes V_j)U_{D_j}$ is an exact $(\alpha_j,n_j,0)$-block encoding of $S_j$. The product construction for block encodings~\cite[Lemma 53, arXiv version]{Gilyen2019} introduces no additional multiplicative normalisation beyond the product of the chosen substage normalisations. Using a fresh $n_j$-qubit ancilla register for each $U_{S_j}$, and repeating this ordered composition over the $T = t/h$ time steps gives 
\begin{equation}
    n_{\rm a} = T\sum_{j=0}^{q_p-1}n_j, \qquad \alpha_\Phi = \left(\prod_{j=0}^{q_p-1}\alpha_j\right)^{T}.
    \label{eq:alpha_phi}
\end{equation}
The required number of ancilla qubits can be reduced exponentially using a compression gadget~\cite[Lemma 3]{Fang2023}, which we demonstrate in the numerical examples later in this work. Equation~\eqref{eq:alpha_phi} shows that when optimally subnormalised block encodings of the dissipative substages are available, there is no additional subnormalisation penalty from the number of stages or steps. In particular, if $\alpha_j =1$, we get an $(\alpha_\Phi,n_{\rm a},0)$-block encoding of $\Phi_p$ with $\alpha_\Phi=1$. If the dissipative substages are instead encoded with $\alpha_j >1$, the success probability contains a subnormalisation penalty that can decay exponentially with the number of splitting steps. In this case, one may be able to further split the operator until a direct block encoding is possible, provided that the dissipative generators remain negative semidefinite. Otherwise, uniform singular-value amplification can reduce this penalty~\cite{Gilyen2019, Fang2023}, but is unnecessary when direct block encodings are available. 

The extent to which the dynamics are non-unitary may be quantified in the continuous and discrete cases by
\begin{equation}
    Q(t) = \frac{\|\vec\psi(0)\|_2}{\|e^{Mt}\vec\psi(0)\|_2}, \quad
    Q_{p,h}(t) = \frac{\|\vec\psi(0)\|_2}{\|\Phi_p(t,h)\vec\psi(0)\|_2}.
    \label{eq:Q_definition}
\end{equation}
Given that $\|e^{Mt}-\Phi_p(t,h)\|\le\epsilon$ and choosing a sufficiently small $\epsilon$ such that $\epsilon Q<1$, they are related by
\begin{equation}
    \frac{Q}{1+\epsilon Q} \le Q_{p,h} \le \frac{Q}{1-\epsilon Q}.
\end{equation}
The discrete state preparation cost therefore converges to that of the physical problem as $Q_{p,h}=Q+O(\epsilon Q^2)$.
The postselection probability is
\begin{equation}
    P_{p,h}(t)=\frac{1}{\alpha_{\Phi}^2Q_{p,h}^2},
    \label{eq:probability}
\end{equation}
using $O(\alpha_\Phi^2Q_{p,h}^2)$ state preparations and block-encoding applications per successful run. This may be improved quadratically to $O(\alpha_\Phi Q_{p,h})$ using amplitude amplification~\cite{Brassard2002} when the block encoding acts on a known input state.

\section{Application to Damped Waves}
\label{sec:application_to_damped_waves}

We use the example from classical physics of a linearly damped wave to demonstrate the approach,
\begin{equation}
    \frac{\partial^2\psi}{\partial t^2} + \gamma\frac{\partial\psi}{\partial t} = c^2\nabla^2\psi,
    \label{eq:damped_wave_equation}
\end{equation}
for damping coefficient $\gamma$ and wave speed $c$. We have chosen an example with a non-trivial quantum circuit derivation to show how the approach can be applied in practice. We use statevector simulations to validate the accuracy, and the IonQ Forte 1 trapped-ion quantum processor~\cite{Chen2024} to assess the feasibility for near-term applications. 

\subsection{Quantum Circuit Derivation}

Applying the Fourier transform to Eq.~\eqref{eq:damped_wave_equation} in one dimension, then adopting a pseudospectral discretisation by $N$ modes corresponding to $N$ discrete grid points, causes each mode to evolve under
\begin{equation}
    \frac{\text{d}^2}{\text{d}t^2}\widehat{\psi}_j + \gamma \frac{\text{d}}{\text{d}t}\widehat{\psi}_j + \omega_j^2\widehat{\psi}_j = 0.
\end{equation}
This is the damped oscillator ODE, where mode $j$ has wavenumber $k_j$ and oscillates with an angular frequency $\omega_j = c|k_j|$ in the undamped case, or $\sqrt{\omega_j^2 -\gamma^2/4}$ in the underdamped case. It can be written as the first-order system
\begin{equation}
    \frac{\text{d}}{\text{d}t}\!\begin{bmatrix}
        \widehat{\psi}_j \\ \text{d}_t\widehat{\psi}_j/\omega_j
    \end{bmatrix}
    =
    \Bigg(
    \underbrace{\begin{bmatrix}
                0 & \omega_j \\
        -\omega_j & 0
    \end{bmatrix}}_{i\omega_jY}
    +
    \underbrace{\begin{bmatrix}
                0 &  0 \\
        0 & -\gamma
    \end{bmatrix}}_{-\gamma\ket{1}\!\bra{1}}
    \Bigg)
    \begin{bmatrix}
        \widehat{\psi}_j \\ \text{d}_t\widehat{\psi}_j/\omega_j
    \end{bmatrix},
    \label{eq:wave_ode}
\end{equation}
where $Y = -i\ket{0}\!\bra{1} + i\ket{1}\!\bra{0}$ is the Pauli-Y gate. The exact evolution is therefore governed by $e^{i\omega_jYt-\gamma\ket{1}\!\bra{1}t}$, which has a closed-form expression with a nonlinear dependence on $\omega_j$, with no readily apparent quantum circuit decomposition. If we instead consider the split dynamics, then $e^{i\omega_jYt} = R_Y^\dagger\!(2\omega_j t)$ and $e^{-\gamma\ket{1}\!\bra{1} t} = \ket{0}\!\bra{0} + e^{-\gamma t}\ket{1}\!\bra{1}$, where $R_Y\!(\theta) = e^{-i\frac{\theta}{2}Y}$ is the standard quantum gate that rotates about the $y$ axis of the Bloch sphere, and $R_Y^\dagger\!(\theta) = R_Y\!(-\theta)$ is its inverse. Hence, operator splitting simplifies the non-trivial dynamics of $e^{i\omega_jYt-\gamma\ket{1}\!\bra{1}t}$ in terms of a standard quantum gate and a simple non-unitary operator that can be readily implemented.

Equation \eqref{eq:wave_ode} describes the evolution of a single mode, while we need a quantum circuit that evolves a statevector containing all modes simultaneously. The quantum amplitudes are prepared either in Fourier space as
\begin{equation}
    \ket{\widehat{\Psi}} \propto \sum_{j=0}^{N-1} \left(\widehat{\psi}_j\ket{0}_S+\frac{\text{d}_t\widehat{\psi}_j}{\omega_j}\ket{1}_S\right)\otimes\ket{j}_D\,,
    \label{eq:spectral_input_state}
\end{equation}
or in physical space as
\begin{equation}
    \ket{\Psi} \propto \sum_{j=0}^{N-1} \left(\psi_j\ket{0}_S+ |c\partial_x|^{-1}\partial_t\psi_j\ket{1}_S\right)\otimes \ket{j}_D,
    \label{eq:physical_space_encoding}
\end{equation}
followed by the quantum Fourier transform (QFT). Since $\omega_0=0$, $\text{d}_t\widehat{\psi}_0/\omega_0$ must be handled separately, and equals $0$ for waves with a velocity that has a zero spatial mean. The initial condition is encoded into the data register~$\ket{\text{dat}}_D$ with the selector qubit~$\ket{\text{sel}}_S$ distinguishing between the wave displacement and the velocity. 

To derive the circuit, we decompose $\omega_j$ using the binary expansion of its index $j = \sum_{r=0}^{n-1} 2^r q_r$, where $q_r\in\{0,1\}$ is a binary digit stored in qubit $r$, $q_0$ is the least-significant qubit, and $N=2^n$. Considering the periodic boundary conditions that are enforced by the QFT, the wavenumber defined on a domain of length $L$ is $k_j = 2\pi j/L$ when $0\leq j < N/2$, and $k_j = (2\pi/L)(j-N)$ when $N/2 \leq j < N$. Therefore, $\omega_j = c|k_j|$ can be written as
\begin{equation}
    \omega_j = \frac{2\pi c}{L}
    \begin{cases}
        \displaystyle\sum_{r=0}^{n-1} 2^r q_r ,       & 0\leq j < 2^{n-1} \\
        2^n - \displaystyle\sum_{r=0}^{n-1} 2^r q_r , & 2^{n-1}\leq j < 2^n .
    \end{cases}
\end{equation}
Substituting the expression into the mode evolution operator gives
\begin{equation}
    R_Y^\dagger\!(2\omega_jt) = 
    \begin{cases}
        R_Y^\dagger\!\left(\zeta t\displaystyle\sum_{r=0}^{n-1} 2^r q_r\right), \!\!&\!\! 0 \leq j < 2^{n-1} \\[1em]
        R_Y^\dagger\!\left(\zeta t\left[2^n {-} \displaystyle\sum_{r=0}^{n-1} 2^r q_r\right]\right), \!\!&\!\! 2^{n-1}\leq j < 2^n,
    \end{cases}
\end{equation}
where $\zeta = 4\pi c/L$ is defined for convenience. Using additivity of rotation angles, this becomes
\begin{equation}
    R_Y^\dagger\!(2\omega_jt) {=} 
    \begin{cases}
        \displaystyle\prod_{r=0}^{n-1} R_Y^\dagger\!(2^r\zeta t q_r), \!\!&\!\! 0 \leq j < 2^{n-1} \\[1em]
        R_Y^\dagger\!(2^n\zeta t)\displaystyle\prod_{r=0}^{n-1} R_Y^\dagger\!(-2^r \zeta t  q_r), \!\!&\!\! 2^{n-1} \leq j < 2^n\,,
    \end{cases}
    \label{eq:wave_gates}
\end{equation}
which is in terms of $R_Y^\dagger$ gates acting on the $\ket{\text{sel}}_S$ register, controlled by qubits in the $\ket{\text{dat}}_D$ register specified by $q_r$.

\begin{figure*}[p]
    \centering
    \includegraphics[width=0.99\textwidth]{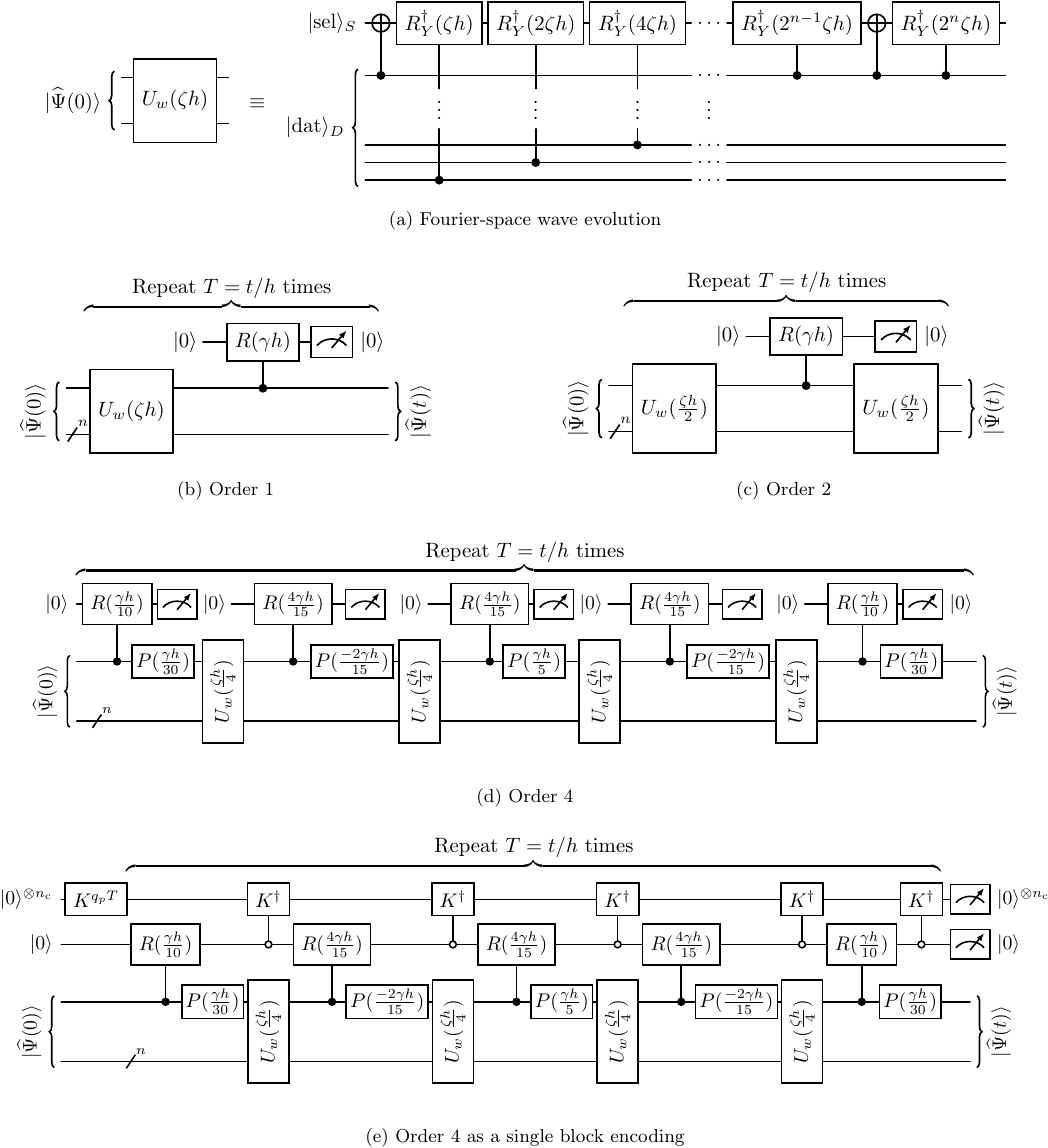}
    \caption{(a) Quantum circuit for solving the wave equation in Fourier space using the initial state in Eq.~\eqref{eq:spectral_input_state}, where $\zeta = 4\pi c/L$. (b--d) Quantum circuits for simulating the damped wave equation in Fourier space with orders 1, 2 and 4. The measurements should be interpreted as acting on separate qubits and all measured at the end of the computation, although mid-circuit measurements of a single qubit are also valid. (e) Compression gadget~\cite{Fang2023} to exponentially reduce the required number of ancilla qubits.  $P(\theta) = \ket{0}\!\bra{0} + e^{i\theta}\ket{1}\!\bra{1}$ is the phase gate, and $R(\theta)$ is an $R_Y\!(\theta)$ gate with a modified argument defined in Eq.~\eqref{eq:R_gate}. The number of counter-register qubits $n_{\rm c}$ is given in Eq.~\eqref{eq:counter_qubits}. The circuits use big-endian ordering.}
    \label{fig:splitting_circuits}
\end{figure*}

The piecewise dependence can be implemented by conditioning each operation on the most-significant qubit $q_{n-1}$, which is $\ket{0}$ for $0\leq j<2^{n-1}$ and $\ket{1}$ for $2^{n-1}\leq j < 2^n$. Since both expressions in Eq.~\eqref{eq:wave_gates} contain the common product term, $R_Y^\dagger\!(\pm 2^r\zeta t q_r)$ up to the sign of the angle, they can be implemented with a positive sign across all $j$ indices, then surrounded by CNOT gates to reverse the sign for the $q_{n-1}=\ket{1}$ component. This avoids the need for additional zero- and one-controlling operations, so all gates are controlled by at most one qubit. Finally, the remaining gate of $R_Y^\dagger\!(2^n\zeta t)$ can be applied to $2^{n-1} \leq j < 2^n$ by controlling on $q_{n-1}=\ket{1}$. The quantum circuit that implements the evolution is shown in Fig.~\hyperref[fig:splitting_circuits]{\ref*{fig:splitting_circuits}a}. 

The damping evolution $\ket{0}\!\bra{0} + e^{-\gamma t}\ket{1}\!\bra{1}$ applies an exponential decay to the velocity component of the state uniformly across all wavenumbers. It can be implemented exactly by applying
\begin{align}
    R(\gamma t) &= R_Y\!(2\arccos [e^{-\gamma t}]) \nonumber \\
    &=
    \begin{bmatrix}
        e^{-\gamma t} & -\sqrt{1 - e^{-2\gamma t}}\\
        \sqrt{1 - e^{-2\gamma t}} & e^{-\gamma t}
    \end{bmatrix}
    \label{eq:R_gate}
\end{align}
to an ancilla qubit in a computational basis state, either $\ket{0}$ or $\ket{1}$, and controlled by $\ket{\text{sel}}_S$. Postselecting the ancilla in its initial basis state implements the desired contraction.

The quantum circuit that simulates the order-1 splitting step in Eq.~\eqref{eq:lie_trotter} is shown in Fig.~\hyperref[fig:splitting_circuits]{\ref*{fig:splitting_circuits}b}, which implements the circuits for wave propagation and damping sequentially. The quantum circuit for the order-2 Strang product formula is shown in Fig.~\hyperref[fig:splitting_circuits]{\ref*{fig:splitting_circuits}c}, where the damping is evolved for time $h$ between wave evolutions for time $h/2$. This is the orientation in Eq.~\eqref{eq:strang} and is used throughout the order-2 simulations and hardware experiments below. The quantum circuit for order 4~\cite{Castella2009} using the product formula in Table \ref{tab:splitting_coefficients} is shown in Fig.~\hyperref[fig:splitting_circuits]{\ref*{fig:splitting_circuits}d}. The ancilla qubits in the presented circuit should be interpreted as different qubits that are all measured at the end of the computation. Interpreting the ancilla wires as the same qubit subject to mid-circuit measurements is also mathematically valid, but a dominant source of error in real hardware~\cite{Gaebler2021}. The circuits for implementing high-order operator splitting use the same constructions of the wave propagation and the real-time damping. The imaginary-time damping $e^{iH_1 \Im(b_j)h} = \ket{0}\!\bra{0} + e^{-i\gamma \Im(b_j)h}\ket{1}\!\bra{1}$ is unitary, corresponding to the phase gate $P(\theta) = \ket{0}\!\bra{0} + e^{i\theta}\ket{1}\!\bra{1}$ with $\theta=-\gamma\Im(b_j)h$ applied to the selector qubit. The same circuit structure in Fig.~\hyperref[fig:splitting_circuits]{\ref*{fig:splitting_circuits}d} applies to splitting of orders 6 and 8, which we have not shown explicitly.

The circuits in Fig.~\hyperref[fig:splitting_circuits]{\ref*{fig:splitting_circuits}b--d} realise the block encoding by postselecting one ancilla qubit for each non-unitary substage, so the high-order product formulas require $q_pT$ ancilla qubits. This construction minimises the required circuit depth per shot, but requires a large number of ancilla qubits. This ancilla requirement can be reduced exponentially using the compression gadget~\cite{Fang2023}, shown in Fig.~\hyperref[fig:splitting_circuits]{\ref*{fig:splitting_circuits}e}. The unitary $K$ is a permutation operator
\begin{equation}
    K = \ket{0}\!\bra{2^{n_\text{c}}-1} + \sum_{j=0}^{2^{n_\text{c}}-2} \ket{j+1}\!\bra{j},
\end{equation}
where
\begin{equation}
    n_\text{c} = \left\lceil \log_2 \left(q_pT\right) \right\rceil + 1
    \label{eq:counter_qubits}
\end{equation}
is the number of qubits in the counter register, which is required in addition to the registers in Fig.~\hyperref[fig:splitting_circuits]{\ref*{fig:splitting_circuits}b--d}.

\begin{figure*}[t]
    \centering
    \includegraphics[width=0.99\textwidth]{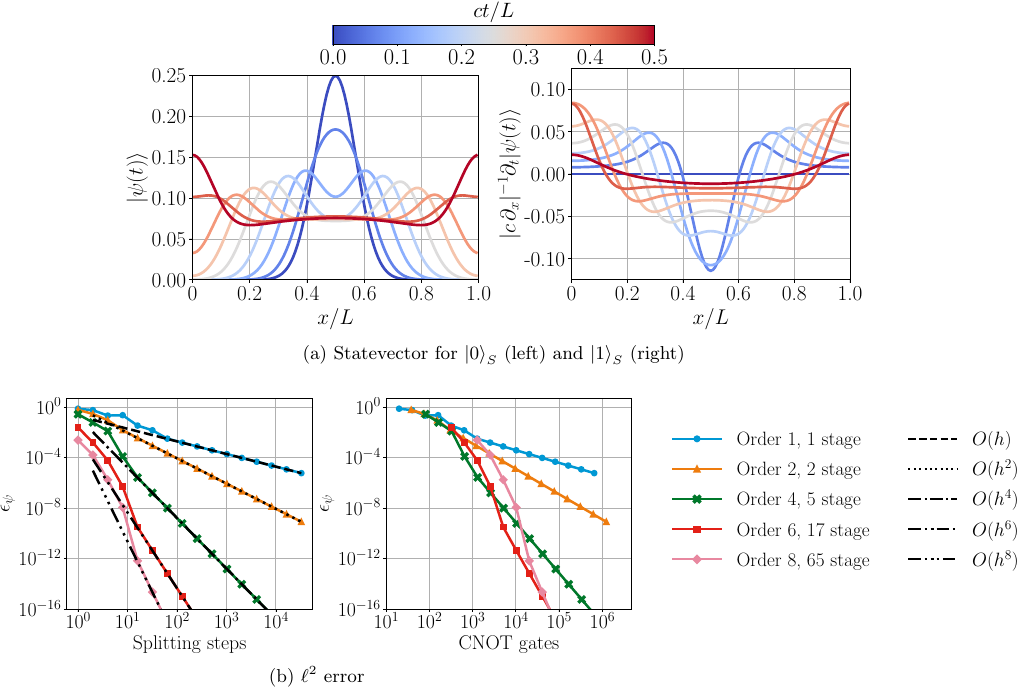}
    \caption{Statevector simulations of the damped wave equation evolution. (a) Physical-space quantities corresponding to the QFT$^\dagger$ applied to the $\ket{\text{dat}}_D$ register using 16 qubits and eight order-4 splitting steps. (b) $\ell^2$ error defined in Eq.~\eqref{eq:l2_error} for all schemes considered against the number of splitting steps (left) and the number of logical CNOT gates in the split evolution, including the compression-counter circuitry (right).}
    \label{fig:statevector_simulations}
\end{figure*}

In $d$ dimensions, the wave modes have angular frequency $c\|\vec{k}\|$. A suitable multidimensional first-order Hamiltonian formulation instead couples the spatial components, and scalable quantum circuits for this evolution have been constructed by \citeauthor{Sato2024}~\cite{Sato2024}. The resulting circuit can be used for each unitary substage $e^{iH_2a_jh}$ of the present splitting method.

The circuits can be extended to homogeneous Dirichlet or Neumann boundary conditions by using the sine or cosine Fourier bases~\cite{Strang1999}, respectively, which also have efficient quantum circuit implementations~\cite{Klappenecker2001}. The wavenumber is then simply $k_j = \pi j/L$ for Neumann conditions or $k_j = \pi (j+1)/L$ for Dirichlet conditions. This was shown by \citeauthor{Pfeffer2025}~\cite{Pfeffer2025} in the context of the advection-diffusion equation in a laminar shear flow, by implementing Neumann boundary conditions with the quantum cosine transform corresponding to a thermally insulated wall.

\subsection{Statevector Simulations}

Figure~\ref{fig:statevector_simulations} presents quantum statevector simulations of the block encoding algorithm, including the compression gadget in Fig.~\hyperref[fig:splitting_circuits]{\ref*{fig:splitting_circuits}e}, applied to the damped wave equation on a periodic domain of length $L$. We consider a Gaussian initial condition of $\psi(x,0) = \exp(-100[x/L-0.5]^2)$ with a zero initial time derivative so $|c\partial_x|^{-1}\partial_t\psi(x,0)=0$. The domain is discretised with 128 grid points, and we assume access to a quantum state encoding this function in the form of Eq.~\eqref{eq:physical_space_encoding}. This requires $n=7$ qubits in the data register, one selector qubit, one block-encoding ancilla qubit, and $n_{\rm c} = \lceil\log_2 q_pT\rceil + 1$ counter-register qubits for the high-order simulations. Figure~\hyperref[fig:statevector_simulations]{\ref*{fig:statevector_simulations}a} uses $p=4$, $q_p=5$, and $T=8$, giving $n_{\rm c} = 7$ counter-register qubits, for 16 total qubits. In Figure~\hyperref[fig:statevector_simulations]{\ref*{fig:statevector_simulations}b}, $q_p$ and $T$ vary, so each simulation uses a different number of qubits. The ratio of the wave propagation to damping time scales is $\gamma L/c=2\pi$, which corresponds to a fundamental-mode damping ratio $\gamma/(2\omega_1)=0.5$, where $\omega_1=2\pi c/L$. In Fig.~\hyperref[fig:statevector_simulations]{\ref*{fig:statevector_simulations}a}, the Fourier-space implementation of each splitting step requires 152 logical CNOT gates, giving 1216 CNOT gates over the full evolution, including the counter-register circuitry. This calculation and its scaling will be expanded upon in the following Section~\ref{sec:resource_requirements}. The probability of successfully postselecting the block encoding by measuring $\ket{0}^{\otimes {n_{\rm c} + 1}}$ at the end of the computation is 27.6\%, corresponding to Eq.~\eqref{eq:probability}. For this initial condition, the steady-state solution where all modes but zero have decayed corresponds to a success probability of $25.1\%$. 

The exact propagator for the Fourier modes can be obtained by evaluating $e^{i\omega_jYt-\gamma\ket{1}\!\bra{1}t}$ classically, which we use to validate the theoretical orders of accuracy with the state $\ell^2$ error, shown later in Eq.~\eqref{eq:l2_error}. The statevector simulations in Fig.~\hyperref[fig:statevector_simulations]{\ref*{fig:statevector_simulations}a}, which use eight order-4 splitting steps, have an $\ell^2$ error of $\epsilon_\psi = 1.26{\times}{10}^{-4}$. This error is plotted in Fig.~\hyperref[fig:statevector_simulations]{\ref*{fig:statevector_simulations}b} for all five product formulas considered here, of orders 1, 2, 4, 6 and 8, against the number of splitting steps and the number of CNOT gates. The results show that all product formulas achieve the theoretical error scaling after the initial coarse-grid transients have passed. The number of logical CNOT gates is comparable at large error tolerances for orders 1, 2, 4 and 6, but the higher-order product formulas rapidly become more efficient for small error tolerances. This suggests that there is little disadvantage in using higher-order product formulas up to order 6, even when low-accuracy solutions are permissible. The order-6 splitting achieves $\epsilon_\psi=10^{-16}$ on the order of $10^5$ CNOT gates. In comparison to the extrapolated costs for the low-order formulas of orders 1 and 2, this is an improvement by approximately 12 and 5 orders of magnitude for the same tolerance. The exponential stage growth in the order limits the practical utility of the order-8 scheme, except in the case of extremely small error tolerances. 

\subsection{End-to-End Hardware Execution}

\begin{figure*}
    \centering
    \includegraphics[width=0.99\textwidth]{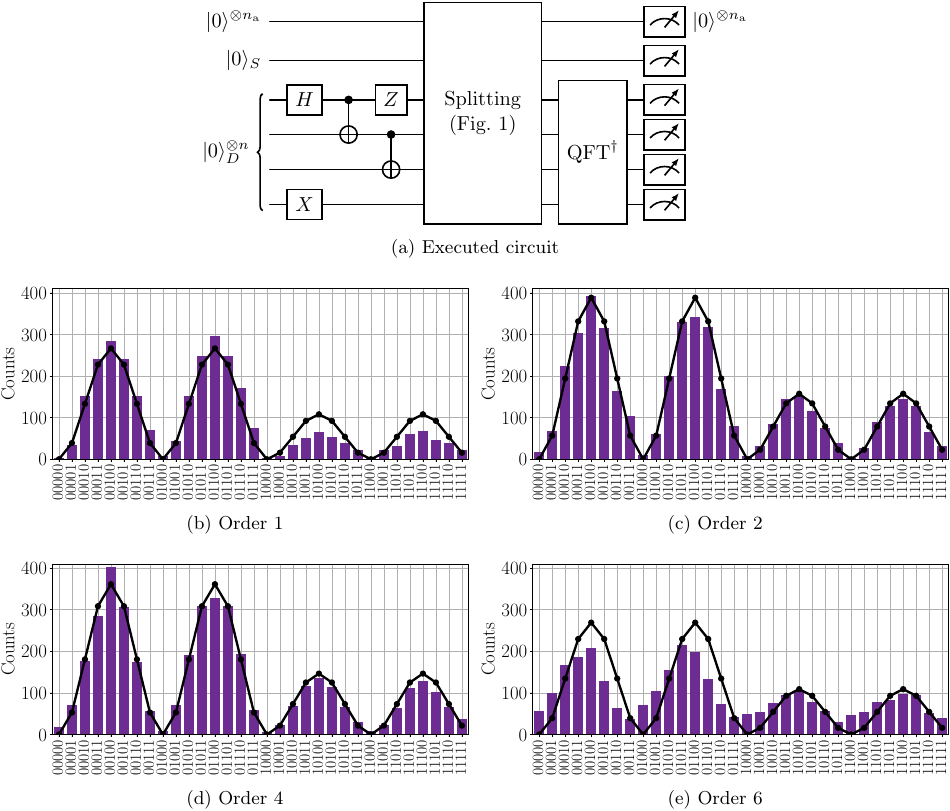}
    \caption{(a) End-to-end quantum circuit for performing a single splitting step with the initial condition in Eq.~\eqref{eq:hardware_state}, using the splitting circuits in Fig.~\ref{fig:splitting_circuits}. (b--e) Measurement histograms after 5000 shots from the IonQ Forte 1 for orders 1, 2, 4 and 6 with the exact, unsplit solution shown by the black line, after rescaling to the measured success probability.}
    \label{fig:hardware_execution}
\end{figure*}

We evaluate the developed high-order product formulas on the IonQ Forte 1~\cite{Chen2024}, which has 36 all-to-all connected trapped-ion qubits and was accessed via AWS Braket. For the damped-wave example, the full end-to-end execution must be considered, including state preparation. To minimise the number of gates required for state preparation, we prepare a sparse state directly in Fourier space as 
\begin{equation}
    \ket{\widehat{\Psi}} = \ket{0}_S \otimes \frac{1}{\sqrt{2}}(\ket{1}_D - \ket{N{-}1}_D)\,,
    \label{eq:hardware_state}
\end{equation}
which corresponds to the physical-space amplitudes 
\begin{equation}
    \ket{\Psi} \propto \sum_{j=0}^{N-1}\sin(2\pi j/N)\ket{0}_S \otimes \ket{j}_D\,.
\end{equation}
Therefore, the Fourier modes correspond to an initial displacement described by a periodic sine wave, with a zero initial velocity. This is a standing wave initialised at its maximum displacement. By preparing the state directly in Fourier space, we also avoid the need for the initial QFT. After evolving the state in Fourier space, we then complete one full inverse QFT to demonstrate that QFT implementation is not prohibitive. The end-to-end quantum circuit is shown in Fig.~\hyperref[fig:hardware_execution]{\ref*{fig:hardware_execution}a}. 

The ratio of wave propagation to damping time scales is again taken to be $\gamma L/c = 2\pi$. We use $N=16$ grid points corresponding to $n=4$ data-register qubits, with one selector qubit required to store both the displacement and velocity data. To minimise the circuit depth per shot, we do not apply the compression gadget~\cite{Fang2023}, and perform all measurements at the end of the computation. Therefore, the end-to-end execution of a splitting step for orders 1, 2, 4, 6 and 8 requires a total of 6, 6, 10, 22 and 70 qubits and 34, 46, 78, 246, and 918 CNOT gates, respectively, including state preparation and the final inverse QFT and six CNOT gates to perform the final SWAP operations. Order 8 requires more qubits than are available, so we simulate orders up to 6 only for one splitting step over the non-dimensional time $ct/L = 0.125$, corresponding to $1/8$ of a full cycle. At this point in the cycle, the undamped evolution has a wave displacement and scaled velocity of equal magnitude, so the effects of damping can be easily visualised. 

The histogram of measurement outcomes conditioned on $\ket{0}^{\otimes n_{\rm a}}$ is shown in Fig.~\hyperref[fig:hardware_execution]{\ref*{fig:hardware_execution}b--e} for orders 1, 2, 4 and 6 respectively, along with the expected measurement outcomes from the exact (unsplit) evolution. For all orders of accuracy, the correct qualitative behaviour of squared sine profiles with a greater magnitude of displacement than velocity is obtained. The results for order 1 are visibly inaccurate, while the effects of noise become visible for order 6 due to the increased circuit depth per step. The histograms for orders 2 and 4 both appear to be highly accurate, having found a good compromise between accuracy and circuit depth for this particular configuration.

Each case was simulated for 5000 shots and had an observed probability of successful postselection of 59.98\%, 87.38\%, 81.26\% and 60.50\%. This compares to the expected probabilities from the corresponding statevector simulations of 60.39\%, 88.40\%, 83.38\%, and 83.36\%, respectively. The expected probability of the true solution is 83.36\%, showing that both high-order product formulas achieve high accuracy in the noise-free statevector simulations. The measurements for order 4 are within $2.10$ percentage points of the true probability, compared to $4.02$ percentage points for order 2. Taking the square roots of the normalised histogram counts gives an estimate of the componentwise magnitudes of the state amplitudes. Evaluating the magnitude-only $\ell^2$ error from $|\vec{\phi}|$ gives 0.1634, 0.1166, 0.1138 and 0.3404 for orders 1, 2, 4 and 6, respectively. Therefore, the results for orders 2 and 4 are of comparable accuracy despite the greater susceptibility to noise from the increased circuit depth in the order 4 simulations. For order 6, the effects of noise dominate, resulting in a lower accuracy than even the order-1 simulations. 

\section{Resource Requirements}
\label{sec:resource_requirements}

We now analyse the resources required to simulate evolution under a general dissipative generator $M=H_1+iH_2$. Since the implementations of the unitary and dissipative substages $V_j$ and $D_j$ are case-dependent, we denote their average CNOT-gate requirements by $G_V$ and $G_D$, respectively. We give the resources required to construct the block encoding to error $\epsilon = \|e^{Mt}-\Phi_p(t,h)\|$, followed by those required to prepare the corresponding normalised quantum state to error
\begin{equation}
    \epsilon_\psi = \left\|\frac{e^{Mt}\vec{\psi}(0)}{\|e^{Mt}\vec{\psi}(0)\|_2}-\frac{\Phi_p(t,h)\vec{\psi}(0)}{\|\Phi_p(t,h)\vec{\psi}(0)\|_2}\right\|_2.
    \label{eq:l2_error}
\end{equation}
We then give the exact resources required for the classical damped-wave simulations.

Since the Hermitian and anti-Hermitian components satisfy $\|H_1\|,\|H_2\|\leq\|M\|$, the global error of an order-$p$ product formula satisfies $\epsilon = O\!\left(t\|M\|^{p+1}h^p\right)$ for fixed $p$. It follows that the required number of time steps $T=t/h$, and hence circuit depth, scales as $T = O([\|M\|t]^{1+1/p}\epsilon^{-1/p})$.

For the selected product formulas, $q_p=2^{p-2}+1$, giving $q_2=2$, $q_4=5$, $q_6=17$ and $q_8=65$. We therefore model the stage count as $O(2^p)$, assuming that this remains representative at higher orders where qualifying schemes are not yet known. This scaling is not universal to all splitting schemes and is used here only as a model for the class considered~\cite{McLachlan1995}. A single application of the resulting $(\alpha_\Phi,n_{\rm a},\epsilon)$-block encoding therefore requires
\begin{equation}
    O\!\left( 2^p(\|M\|t)^{1+1/p}\epsilon^{-1/p} [G_V+G_D]\right)
    \label{eq:block_encoding_complexity}
\end{equation}
CNOT gates. Without ancilla compression, the block encoding requires $n_{\rm a} = T\sum_{j=0}^{q_p-1}n_j = O(2^pT)$ ancilla qubits when each dissipative substage uses $O(1)$ ancillas. The compression gadget~\cite{Fang2023} reduces this to
\begin{equation}
    n_{\rm a} = \max_j n_j+\left\lceil\log_2(q_pT)\right\rceil+1,
\end{equation}
at the cost of an additive term such that the CNOT complexity in Eq.~\eqref{eq:block_encoding_complexity} becomes
\begin{equation}
    O\!\left( 2^p[\|M\|t]^{1+1/p}\epsilon^{-1/p} \left[G_V+G_D+\operatorname{polylog}(q_pT)\right] \right).
\end{equation}

We now consider preparation of the normalised output state. We use $Q$ as defined in Eq.~\eqref{eq:Q_definition} to quantify the norm loss along the simulated evolution. Unitary dynamics have $Q=1$, while $Q>1$ indicates contraction of the evolved state. The block-encoding error $\epsilon$ and state error $\epsilon_\psi$ satisfy $\epsilon_\psi \leq 2\epsilon Q$. Therefore, achieving a state error $\epsilon_\psi$ requires a block-encoding error $\epsilon=O(\epsilon_\psi/Q)$. Substituting this into Eq.~\eqref{eq:block_encoding_complexity}, a single block-encoding application at the required accuracy has CNOT complexity
\begin{equation}
    O\!\left(2^p[\|M\|t]^{1+1/p}Q^{1/p}\epsilon_\psi^{-1/p} [G_V+G_D]\right)
\end{equation}
without the compression gadget. Postselecting the ancilla qubits succeeds with probability $P_{p,h}(t) = O(\alpha_\Phi^{-2} Q^{-2})$ as given in Eq.~\eqref{eq:probability}.  Direct postselection therefore requires $O(\alpha_\Phi^2Q^2)$ attempts per successful output state. When the initial state is prepared by a known unitary and a success probability lower bound is available, amplification~\cite{Brassard2002} may be applied to the block encoding, requiring just $O(\alpha_\Phi Q)$ coherent applications of the block encoding and its inverse. The resulting CNOT complexity using amplitude amplification is
\begin{equation}
    O\!\left( 2^p\alpha_\Phi(Q\|M\|t)^{1+1/p}\epsilon_\psi^{-1/p}[G_V+G_D] \right).
\end{equation}
In many dissipative settings, $\alpha_\Phi=1$, as discussed surrounding Eq.~\eqref{eq:alpha_phi}.

Figure~\ref{fig:splitting_circuits} uses the controlled $R_Y\!(\theta)$ gate, which can be factored into
\begin{equation*}
    \left(I\otimes R_Y\!\left[\tfrac{\theta}{2}\right]\right) \operatorname{CNOT} \left(I\otimes R_Y\!\left[-\tfrac{\theta}{2}\right]\right) \operatorname{CNOT},
\end{equation*}
requiring two CNOT gates~\cite{Barenco1995}. For $n$ data-register qubits, the wave evolution $U_w(t)$ requires $2n+4$ CNOT gates, while the damping evolution requires two CNOT gates. Without the compression gadget, high-order splitting with $q_p$ stages therefore requires $T([2n+4][q_p-1]+2q_p)$ CNOT gates in Fourier space. This expression also gives the CNOT depth because every CNOT acts on, or is controlled by, the selector qubit $\ket{\mathrm{sel}}_S$. For the compression gadget, we keep the counter register in the Fourier basis, where the cyclic permutation $K$ is diagonal. Since the initial counter value $q_pT$ is known classically, its Fourier-basis state can be prepared using only single-qubit gates, while the final projection onto zero likewise requires only single-qubit gates~\cite{Draper2000}. Each controlled decrement $K^\dagger$ then consists of $n_{\rm c}$ controlled phase gates, each of which can be decomposed exactly using two CNOT gates~\cite{Barenco1995}. A controlled decrement therefore requires $2n_{\rm c}$ CNOT gates. Since there are $q_pT$ such decrements, the compressed $T$-step evolution requires $T([2n+4][q_p-1]+2q_p[1+n_{\rm c}])$ CNOT gates in Fourier space. For the order-4 simulation in Fig.~\ref{fig:statevector_simulations}, $n=7$, $q_p=5$, $T=8$, and $n_{\rm c}=7$, giving a total of 1216 CNOT gates.

\section{Conclusions}
\label{sec:conclusions}

We have developed a method for constructing block encodings of dissipative evolutions by high-order operator-splitting approximations. The evolution generator is decomposed into Hermitian and anti-Hermitian components, which are typically much simpler to implement than the exact dynamics. The resulting algorithm prepares a block encoding by sequences of simple Hamiltonian evolutions in real and imaginary time, with a composition that achieves high-order accuracy. If each block encoding of the simpler dissipative substages can be optimally subnormalised, their composition introduces no additional subnormalisation penalty in the block encoding of the entire time evolution. The method extends the flexibility and accuracy of high-order splitting methods for simulating unitary dynamics on quantum computers to dissipative dynamics, which are ubiquitous across computational science.

Noise-free statevector simulations confirmed the convergence behaviour for all presented product formulas. For the damped-wave example, orders 1, 2, 4 and 6 had comparable logical CNOT requirements for tolerances of $\epsilon_\psi\approx 10^{-2}$. For tolerances of $\epsilon_\psi\approx 10^{-6}$ and beyond, the high-order constructions reduced the CNOT-gate requirements by at least one order of magnitude, with the advantage widening as the tolerance decreases. This suggests that high-order splitting can be chosen robustly across large error tolerances, and is more efficient at smaller tolerances. The order-8 scheme is less competitive, only becoming useful at extremely small tolerances that are unlikely to be required in practice. 

The quantum circuits were executed on the IonQ Forte~1 trapped-ion quantum processor using product formulas up to order 6. The order-4 approximation reproduced the dynamics more accurately than order 1 and with accuracy comparable to order 2, while also yielding success probabilities closer to the ideal contractive evolution.

For the selected formulas and under the stated substage assumptions, neglecting the polylogarithmic overhead associated with the counter register, the algorithm requires 
\begin{equation*}
O\!\left(2^p\alpha_\Phi(Q\|M\|t)^{1+1/p} \epsilon_\psi^{-1/p}[G_V+G_D] \right)
\end{equation*}
two-qubit gates to prepare an $\epsilon_\psi$-accurate quantum state proportional to the dissipative solution. High-order splitting gives a near-linear dependence on the simulation time arising from time discretisation. For order 6, this is $O(t^{7/6})$, improving substantially over the $O(t^2)$ and $O(t^{3/2})$ for orders 1 and 2. 

The present analysis assumes a time-independent generator $M$. High-order splitting methods for irreversible problems with time-dependent generators $M(t)$ have been developed by augmenting the system with time as an additional coordinate~\cite{Seydaouglu2014}. In the context of quantum algorithms, high-order methods for time-dependent Hamiltonian simulation have also been developed~\cite{Ikeda2023}. Extending the present block-encoding construction to this setting would additionally require efficient high-order implementations of the resulting time-dependent unitary substages, and is left for future work.

The order-8 construction clarifies that higher formal order does not translate indefinitely into lower quantum-circuit costs. Rather than constructing a single product formula with an exponentially growing number of stages, linear combinations of lower-order formulas may offer a more efficient route to faster convergence, as has already been demonstrated for unitary dynamics~\cite{Childs2012, Low2019}. Whether this strategy can deliver similar benefits for non-unitary dynamics is an open question.

Finally, the practicality of the method relies on efficient quantum circuits for the unitary substages and efficient, optimally subnormalised block encodings of the dissipative substages. Developing practical constructions for a broad range of applications is important for advancing industry readiness for fault-tolerant quantum computing.

\bibliography{apssamp}

\section*{Acknowledgements}
Peter Brearley is supported by The University of Manchester via the Dame Kathleen Ollerenshaw Fellowship and the National Quantum Computing Centre [NQCC200921], which is a UKRI Centre and part of the UK National Quantum Technologies Programme (NQTP). Philipp Pfeffer is supported by the European Union (ERC, MesoComp, 101052786). Views and opinions expressed are however those of the author(s) only and do not necessarily reflect those of the European Union or the European Research Council. Neither the European Union nor the granting authority can be held responsible for them.

\appendix
\setcounter{table}{0}
\renewcommand{\thetable}{\Alph{table}}
\renewcommand{\theHtable}{appendix.\Alph{table}}

\begin{table*}[p]
\centering
\caption{The new order-8 product formula with positive real coefficients $a_j=1/64$ for $j=0,\dots,63$, and complex coefficients with positive real part $b_j=b_{64-j}$. Coefficients are reported to 64 significant figures. The coefficients satisfy $\sum_j b_j = 1$ with $\|\mathbf b\|_1 = \sum_j|b_j|=1.124$, $\|\mathbf b\|_2 = \sqrt{\sum_j|b_j|^2}=0.1445$, $\min_j \Re (b_j) \approx 0.007107$ and $\min_j |\Im(b_j)| \approx  0.001529$.}
\label{tab:order8_coefficients}
\footnotesize
\begin{tabular}{@{}r@{\;}c@{\;}l@{}}
\toprule
$b_0 = b_{64}$ & $=$ & $0.007107369987083011266371829075662443421028443768368291174921923276$ \\[-0.4mm]
& $+$ & $0.003834328456660828711273818366899223247961329068041000707659037872\,i$ \\[0.27mm]
$b_1 = b_{63}$ & $=$ & $0.01616425615776470818361036703384222431775264668775198567772662678$ \\[-0.4mm]
& $-$ & $0.01172345949128380894179419684976195966725072071258208620784958855\,i$ \\[0.27mm]
$b_2 = b_{62}$ & $=$ & $0.01005565943341353311473491651971404989964167762044566321146179030$ \\[-0.4mm]
& $+$ & $0.005824421757352811866544912579368939656940058472191123313745250353\,i$ \\[0.27mm]
$b_3 = b_{61}$ & $=$ & $0.02069997864044040783777928609320199559386107186634720278132942821$ \\[-0.4mm]
& $+$ & $0.008274940762536383275230460315586600392288168309913940797608099725\,i$ \\[0.27mm]
$b_4 = b_{60}$ & $=$ & $0.02046191619935861510226444136003630736255753903059489494664182226$ \\[-0.4mm]
& $-$ & $0.001529291827181195766732138888204870863204554960413646121591262140\,i$ \\[0.27mm]
$b_5 = b_{59}$ & $=$ & $0.01835108723669403471748166516094444617092753592478862596384423100$ \\[-0.4mm]
& $-$ & $0.007210245190424043838759073929057672456712276673861278134361785300\,i$ \\[0.27mm]
$b_6 = b_{58}$ & $=$ & $0.009258806102379592469947245114141781117634926380554278376082570234$ \\[-0.4mm]
& $-$ & $0.006788465429204836292820749809815395124074470827015540928183844736\,i$ \\[0.27mm]
$b_7 = b_{57}$ & $=$ & $0.01808445346537030222289275285757692133973681774817376772392892011$ \\[-0.4mm]
& $+$ & $0.009905168332265220483805711793315550354456529293612268562307941074\,i$ \\[0.27mm]
$b_8 = b_{56}$ & $=$ & $0.01191323528273744048855572629265138957199645784884998108526551691$ \\[-0.4mm]
& $+$ & $0.01087892260273590442891538796543162405106265230891411369363451341\,i$ \\[0.27mm]
$b_9 = b_{55}$ & $=$ & $0.01420174952487264547471432791560321162384974799043632555902748179$ \\[-0.4mm]
& $-$ & $0.008473057085889524163755449869854302780620800974268540696154298275\,i$ \\[0.27mm]
$b_{10} = b_{54}$ & $=$ & $0.01050351013568487900105625320264701625835270708256800139971672601$ \\[-0.4mm]
& $+$ & $0.002941900023841630643377739671103431764158759992775427952440336674\,i$ \\[0.27mm]
$b_{11} = b_{53}$ & $=$ & $0.01726549670025747519569869210302584654996040341550147865353036718$ \\[-0.4mm]
& $-$ & $0.01118893898334276896963374679044674522178749410318641467062967763\,i$ \\[0.27mm]
$b_{12} = b_{52}$ & $=$ & $0.01483826263459813245780906238502400184809239620088652286409899470$ \\[-0.4mm]
& $+$ & $0.001973788582738519070001791891024330890863592705817353737129126887\,i$ \\[0.27mm]
$b_{13} = b_{51}$ & $=$ & $0.02222412521741256897868123767718642597092099833964057320284135380$ \\[-0.4mm]
& $-$ & $0.007503457389752695009357880927630719280145722329550746508846129150\,i$ \\[0.27mm]
$b_{14} = b_{50}$ & $=$ & $0.02112299320194063525518767157693887134059987235232710457261008363$ \\[-0.4mm]
& $+$ & $0.005043660757831455485108788840911524134157904299310422514055270445\,i$ \\[0.27mm]
$b_{15} = b_{49}$ & $=$ & $0.01725124024594259866414965164655132062299118205522689660460763958$ \\[-0.4mm]
& $+$ & $0.002448354754246169229300432850269801059070985247540446598662580805\,i$ \\[0.27mm]
$b_{16} = b_{48}$ & $=$ & $0.01720164696443782139524629960119988323515035107670797949802231996$ \\[-0.4mm]
& $+$ & $0.009832763032936267588363321025763445800529250093788287282319236070\,i$ \\[0.27mm]
$b_{17} = b_{47}$ & $=$ & $0.008489616907619475041190849310436165245284310691808874272830986259$ \\[-0.4mm]
& $+$ & $0.005532417787928327727290911327327662726210241736617339262601081905\,i$ \\[0.27mm]
$b_{18} = b_{46}$ & $=$ & $0.008934501759096144531314490101755112492639182871022266824575562959$ \\[-0.4mm]
& $-$ & $0.009857403391643407077646095428223203487949564934060018927167817613\,i$ \\[0.27mm]
$b_{19} = b_{45}$ & $=$ & $0.01889125888840681089548921348216938645332003256315448324868805023$ \\[-0.4mm]
& $-$ & $0.01246520895226434923866589367680779792790948217452623840263778325\,i$ \\[0.27mm]
$b_{20} = b_{44}$ & $=$ & $0.01629697707167399706692590109200154168161270621587330782909411306$ \\[-0.4mm]
& $+$ & $0.01240234374456374062506595192779268958286265059206866886825147492\,i$ \\[0.27mm]
$b_{21} = b_{43}$ & $=$ & $0.01375557172033695698999131068477357548622287310631461283850325199$ \\[-0.4mm]
& $-$ & $0.006485476544806521506247518957432042628245665266406321509450695824\,i$ \\[0.27mm]
$b_{22} = b_{42}$ & $=$ & $0.01685177583349167562540203722611833800946650193875610488961391638$ \\[-0.4mm]
& $+$ & $0.008205683362987041996856504800083148482430873998151826323375036753\,i$ \\[0.27mm]
$b_{23} = b_{41}$ & $=$ & $0.01569231868631648181792320657571675654896406573737063079214773953$ \\[-0.4mm]
& $-$ & $0.003077442692383647718459717456112269598957823156468725484017778338\,i$ \\[0.27mm]
$b_{24} = b_{40}$ & $=$ & $0.01275160419977394639353203781515717356200725356586901064600859808$ \\[-0.4mm]
& $+$ & $0.001639833025847390267544683812978398875142837780668614634148975972\,i$ \\[0.27mm]
$b_{25} = b_{39}$ & $=$ & $0.02357734254570492891786438478918645955394051754339384716392187823$ \\[-0.4mm]
& $-$ & $0.002981205186470358996856901675879127768403542650108637214956237035\,i$ \\[0.27mm]
$b_{26} = b_{38}$ & $=$ & $0.01243037413172166437979887818375860209919471544488816012942955342$ \\[-0.4mm]
& $+$ & $0.008070169921614642158146362417904591999319416699763579562348623357\,i$ \\[0.27mm]
$b_{27} = b_{37}$ & $=$ & $0.01970082122199184297080523084502316151699432918270956935029787593$ \\[-0.4mm]
& $-$ & $0.009727609645674796729980667433904113938444391391011792248285798853\,i$ \\[0.27mm]
$b_{28} = b_{36}$ & $=$ & $0.01235217250734736520376547760151859946867457001233773143576894146$ \\[-0.4mm]
& $-$ & $0.002133159251026507101709119866378423070567491002056507205013899838\,i$ \\[0.27mm]
$b_{29} = b_{35}$ & $=$ & $0.01365582029433676454623773295113766885990943660743360436071112990$ \\[-0.4mm]
& $-$ & $0.001554077063581551839693183226616567757020178147324132766449051360\,i$ \\[0.27mm]
$b_{30} = b_{34}$ & $=$ & $0.02237942354135703633959606016434480604304414856661381293782469946$ \\[-0.4mm]
& $+$ & $0.008512150738243683636631557184571878185732724614193194507613430872\,i$ \\[0.27mm]
$b_{31} = b_{33}$ & $=$ & $0.007592323650719984229208212146106157179261984654742622446562320645$ \\[-0.4mm]
& $+$ & $0.008197988413831796186824177749449465737623921888757577954380597536\,i$ \\[0.27mm]
$b_{32}$ & $=$ & $0.01988461981943304644954710282969671910881719181708357507672717346$ \\[-0.4mm]
& $-$ & $0.02164067586646360037634035946731419073903543559856911849336993348\,i$ \\
\bottomrule
\end{tabular}
\end{table*}

\end{document}